\documentclass[apj]{emulateapj}
\usepackage{array} 
\usepackage{amsmath}
\usepackage{wasysym}

\newenvironment{tightcenter}{%
	\setlength\topsep{0pt}
	\setlength\parskip{0pt}
	\begin{center}
}{%
 	\end{center}
}

\newcommand\chandra{{\it Chandra}}
\newcommand\GOES{{\it GOES}}
\newcommand\ciao{{\it CIAO}}

\newcommand\caldb{{\it CALDB}}

\shorttitle{Cometary Dust/Ice Scattering Emissions}
\shortauthors{Snios et al.}

\begin{document}

\title{The Presence of Dust and Ice Scattering in X-ray Emissions from Comets}

\author{Bradford Snios$^{1}$} 
\author{Jack Lichtman$^{2}$}
\author{Vasili Kharchenko$^{1,2}$}
\affil{${}^{1}$Harvard-Smithsonian Center for Astrophysics, 60 Garden Street, Cambridge, MA 02138 USA}
\affil{${}^{2}$Department of Physics, University of Connecticut, Storrs, CT 06269, USA}

\begin{abstract}
X-ray emissions from cometary atmospheres were modeled from first principles using the charge-exchange interaction with solar wind ions as well as coherent scattering of solar X-rays from dust and ice grains. Scattering cross-sections were interpolated over the 1 nm--1 cm grain radius range using approximations based on the optically thin or thick nature of grains with different sizes. The theoretical emission model was compared to \chandra{} observations of Comets ISON and Ikeya--Zhang due to their high signal-to-noise ratios and clearly defined spectral features. Comparing the observed intensities to the model showed that the charge-exchange mechanism accurately reproduced the emission spectra below 1 keV, while dust and ice scattering was negligible. Examining the 1--2 keV range found dust and ice scattering emissions to agree well with observations, while charge-exchange contributions were insignificant. Spectral features between the scattering model and observations also trended similarly over the 1--2 keV range. The dust and ice density within the cometary atmosphere $n$ was varied with respect to grain size $a$ as the function $n(a) \propto a^{-\alpha}$, with Ikeya--Zhang requiring $\alpha = 2.5$ and ISON requiring $\alpha = 2.2$ to best fit the observed spectral intensities. These grain size dependencies agreed with independent observations and simulations of such systems. The overall findings demonstrate evidence of significant scattering emissions present above 1 keV in the analyzed cometary emission spectra and that the dust/ice density dependence on grain radius $a$ may vary significantly between comets.  
\end{abstract}


\section{Introduction}
\label{Intro}

Cometary X-ray emissions were originally discovered by \cite{Lisse1996} from Comet C/1996 B2 (Hyakutake) and are now observed in over 30 comets. As a comet passes through the Solar System, it absorbs solar energy and ejects neutral material from its surface outward with velocities of a few km s$^{-1}$ via sublimation and localized jet-streams \citep{Wegmann2004, AHearn2011}. The neutral ejecta forms a diffuse cometary atmosphere that undergoes significant transformation from photo-chemical reactions, the dissociation of molecular species, and the fragmentation of dust and ice particles. The neutral particles that compose the cometary atmosphere generate X-rays from the charge-exchange (CX) interaction with solar wind (SW) ions as well as through fluorescence and coherent scattering of solar X-rays from dust and ice grains \citep{Cravens1997, Krasnopolsky1997, Kharchenko2003, Lisse2004b, Bodewits2007, Dennerl2010, Ewing2013, Snios2016}.  

CX emissions are generated from collisions between highly-charged, heavy SW ions ($\sim$0.1\% of all SW ions) and the neutral cometary gas, and CX is known to be the dominate emission mechanism from comets \citep{Krasnopolsky1997, Kharchenko2003, Lisse2004b}. While cometary CX emission is well documented at energies less than 1 keV, \chandra{} observations have shown spectral features for Comet Ikeya--Zhang at energies between 1 and 2 keV \citep{Ewing2013}. The current interpretation of this observed spectrum is that the hard X-ray peaks are a result of the CX from the abnormal, highly ionized SW ions Mg$^{11+}$ and Si$^{13+}$ \citep{Bodewits2007,Ewing2013}. However, in situ observations of the SW using mass spectrometers have never detected these highly-charged ions \citep{vonSteiger2000,Lepri2013}. Theoretical SW plasma models also predict an infinitesimally low probability of finding Mg$^{11+}$ and Si$^{13+}$ in the SW plasma due to the inability to reach such high freezing-in temperatures at regular SW and coronal mass ejection temperatures \citep{Bochsler2007}.

Given the issues using CX to induce X-ray emissions above 1 keV, recent analyses by \cite{Snios2014} and \cite{Snios2016} have explored the possibility of alternative emission mechanisms being responsible for these hard X-ray features. Solar X-ray emissions above 1 keV have been observed to increase by 1--3 orders of magnitude during solar flares, which would also increase scattering contributions by cometary neutral particles by the same magnitude \citep{Neupert2006, Snios2014}. Such an increase may result in scattering, providing a significant contribution of cometary X-ray emissions, and potentially equalling the spectral intensities observed between 1--2 keV. Furthermore, an analysis of observed disk X-ray emissions from Jupiter found that Jovian disk spectra possess similar intensity peaks at energies greater than 1 keV to those seen from comets, and it is known that the primary X-ray production mechanism from the Jovian disk is the scattering of solar X-rays  \citep{BranduardiRaymont2007, BranduardiRaymont2008}. It is therefore probable that scattering of solar X-ray photons by cometary dust and ice grains is the most likely alternative emission candidate.

Significant research has previously been performed to include dust scattering in cometary emission spectra, though no conclusive evidence of X-ray scattering emissions has yet been shown \citep{Owens1998, Wickramasinghe2000, Schulz2006}. Such a detection would provide valuable insight on properties like densities, cross-sections, and grain size distribution of the dust and ice particles present within these systems. Probing dust and ice through X-rays will also let us investigate nano-particles contributions, of which little is understood due to a lack of detector sensitivity at such small grain radii  from in situ observations \citep{UtterbackKissel1990, Rotundi2015}. These findings would not only be applicable to comets but also to any diffuse neutral systems, such as outflows from planetary atmospheres or halo dust emissions from stars. 

In this article, we aim to further quantify dust and ice scattering emissions from cometary atmospheres and investigate its emission strength relative to CX. To do so, the scattering research of \cite{Snios2014} was expanded to include average emissions from all dust and ice grain sizes. The CX model from \cite{Snios2016} was also included, and the total emission model was used to find the best fit for the observational cometary data. Total contributions from each mechanism were found for different spectral intervals and were identified at the energies for which each mechanism was dominant. As comets with emission features greater than 1 keV were required to test our hypothesis, the sample size of available comets was small due to low signal-to-noise ratios for an average observation (J. Lichtman et al. 2018, submitted). Despite the limitations, the results find evidence of significant scattering emissions present above 1 keV in the analyzed emission spectra. 

\section{Modeling Comet X-Ray Emissions}
\label{sect:model}

To model the total emission from a cometary atmosphere, we elected to focus on CX and scattering as they are the most significant emission mechanisms over the observed energy range \citep{Krasnopolsky1997, Snios2014}.  The fluorescence mechanism may be also be important for emissions above 1 keV if the dust particles are primarily composed from silicate materials, such as olivine. Si, Mg, and Fe atoms can also provide fluorescent photons above 1 keV. However, cometary atmospheres are primarily water ice particles and carbon-based dust \citep{Lisse2005, Biver2006, Christian2010}, both of which emit fluorescent photons with energies below 1 keV, an energy range dominated by CX.  Our consideration therefore ignored fluorescence, although the resonance fluorescence of heavier elements can be introduced using K-shell absorption cross-sections of heavier elements \citep{Snios2014}.

CX emissions were modeled with the work outlined in \cite{Snios2016}, which utilized SW composition ratios, physical properties of the cometary atmosphere, and the observation geometry to produce a CX spectrum from first principles. The CX contributions from heavy SW ions that have been detected via in situ observations were included, which consisted of the following: C$^{5+}$, C$^{6+}$, N$^{5+}$, N$^{6+}$, N$^{7+}$, O$^{6+}$, O$^{7+}$, O$^{8+}$, Ne$^{8+}$, Ne$^{9+}$, Mg$^{9+}$, Mg$^{10+}$, Si$^{10+}$, S$^{9+}$, S$^{10+}$, S$^{11+}$, Fe$^{10+}$, Fe$^{11+}$, Fe$^{12+}$, and Fe$^{13+}$. The SW composition may then be varied until a best fit with observational data is achieved via $\chi^{2}$ minimization. 

Scattered emissions were modeled using the work discussed in  \cite{Snios2014}. Given that the maximum radius of the cometary atmosphere is generally 2--3 orders of magnitude smaller than the comet--Sun and comet--detector distances for an average observation, the spectral intensity $I_{sc}(\epsilon)$ of X-rays scattered by the cometary atmosphere for $j$ types of scatterers can be described by the simplified formula
\begin{equation}
	\label{I_total}
	I_{sc}(\epsilon)  = I_{0}(\epsilon) \frac{R_{0}^2}{r_{c}^{2} \Delta^2} \sum\limits_{j} \int{n_{j}(\vec{r}, a) \sigma_{j}(a, \theta_{sc}, \epsilon) dV da}, 
\end{equation} 
 where $I_{0}(\epsilon)$ is the observed solar X-ray intensity at the detector, $R_{0}$ is the Sun--detector distance, $r_c$ is the comet--Sun distance,  $\Delta$ is the comet--detector distance, $n(\vec{r},a)$ is the dust/ice particle density within the atmosphere, $\vec{r}$ is the cometocentric radius vector, $a$ is the grain size, $\sigma$ is the scattering cross-section, and $\theta_{sc}$ is the scattering angle. \cite{FinkRubin2012} defined $n(r,a)$ as 
\begin{equation}
	n(\vec{r},a) \propto \frac{a^{-\alpha}}{r^{\beta}} ,
	\label{n_dust_new}
\end{equation} 
where $\alpha$ is set to 2.5 and $\beta$ is set to 2.0. This distribution agrees with the interpolated results from models \citep{Rubin2011} and with observations \citep{UtterbackKissel1990, Rotundi2015}. Despite these agreements, $\alpha$ is dependent on various factors of the system including comet size, jet stream presence, jet stream locations, comet--Sun distance, comet composition, and comet origin \citep{UtterbackKissel1990, Rubin2011, FinkRubin2012}. We therefore allowed $\alpha$ to vary within this physical range in order to find the best fit to the observational data, while $\beta$ was left fixed at 2.0.

To quantify total scattered emissions, we considered possible contributions from dust grains of all sizes. The upper limit of the grain radius was set to 1 cm based on observations from \cite{UtterbackKissel1990} and \cite{Rotundi2015}, and a lower limit of 1 nm was selected based on theoretical work of \cite{Snios2014}. The mass-loss rate was held fixed and distributed over the wider range of particles to ensure that the physical constraints of the system were preserved. By considering scattered emissions from all grain sizes, cross-sections must be determined to a reasonable accuracy over the entire radius range. The Mie scattering model, used in previous emission modeling \citep{Snios2014, Lewkow2016}, is not applicable for nano-sized grains, which require quantum mechanical calculations. It is also not valid for the large-size grains with stochastic shapes, structures, and high levels of  porosity. Rather than derive the cross-section for every grain size,  shape, and porosity, an approximate relationship to describe cross-section as a function of grain radius can be developed. 

\begin{table*}
	\caption{\chandra{} Comet Observation Parameters}
	\label{table:parameter} 
	\begin{tightcenter}
	\begin{tabular}{ c c c c c c c c c c c}
		\hline \hline
		& & & $T_{\rm exp}$ & $r_{c}$ & $\Delta$ & $Lat_{\astrosun}$ 
			& $Long_{\astrosun}$ & $Q_{H_{2}O}$ & $\cal{I}_{\rm GOES,\ 4-8 \AA}$ \\
		Comet & Prop. Num. & Obs. Date &  (ks) & (au) & (au) & (degree) 
			& (degree) & (10$^{28}$ mol s$^{-1}$) &  $(10^5 \rm\ photons\ cm^{-2}\ s^{-1})$ \\
		\hline
		IZ  & 03108076 & 2002 April 15--16 & 24 & 0.81 
			& 0.45 & 206.7 & 26.49 & 20\tablenote{\cite{Biver2006}} & 30 \\
		ISON & 15100583 & 2013 Oct 31--Nov 6 & 36 & 1.18 & 0.95 
			& 1.130 & 115.0 & 2\tablenote{\cite{Combi2014b} }& 12 \\
		\hline
	\end{tabular}
	\end{tightcenter}
\end{table*}

To begin, consider a porous, optically thin dust grain that will approximate a small grain particle. Assuming that summation over all possible grain configurations will produce an isotropic shape as its average, the total number of particles present in the grain is
\begin{equation} 
	N_{g}(a) = \frac{4\pi}{3}a^3 n_{p} ,
\end{equation}
where $n_{p}$ is the atomic, or molecular, particle density within the grain. If the particle is porous enough where each atom or molecule may be considered non-interacting with any other, the total cross-section of the small-size grain $\Sigma_{thin}$ equals 
\begin{equation} 
	\label{sigma_thin}
	\Sigma_{thin}(a) = N_{g}(a) \sigma_{p} = \frac{4\pi}{3}a^3 n_{p} \sigma_{p} ,
\end{equation}
where $\sigma_{p}$ is the cross-section of an individual atom or molecule. The obtained Equation~\ref{sigma_thin} describes optically thin dust grains but also can be applied to optically thin gas objects, and a comparison to the results found from Mie scattering for spherical, nano-sized particles showed strong agreement \citep{vandeHulst1981, Draine2003}. Equation~\ref{sigma_thin} was therefore used as the small cross-section approximation.  

To calculate the upper limit of the cross-section range, recall that for a macroscopic spherical particle that completely absorbs radiation, the total cross-section is defined as 
\begin{equation} 
	\label{sigma_thick}
	\Sigma_{thick}(a) = 2\pi a^2 ,
\end{equation}
which can be treated as the upper limit of the total cross-section for optically thick grains. The factor 2 in Equation~\ref{sigma_thick} reflects the fact that total cross-sections are a sum of the absorption (geometrical) cross-section $\pi a^{2}$ and the diffraction scattering cross-sections at the limit  $\lambda/a \rightarrow 0$, where $\lambda$ is the wavelength of the absorbed radiation \citep{Landau1958}. Eqs.~\ref{sigma_thin} and \ref{sigma_thick} provide two physical limits of cross-sections for optically thin and thick grains, assuming grain size $a$ can be considered a variable physical parameter. A simple interpolation formula for the grain total cross-section $\Sigma_{total}$ can be constructed that correctly describes cross-section behavior at small and large grain sizes.  The suggested cross-section interpolation should therefore evolve over this grain radius range as
\begin{equation} 
	\label{sigma_total}
	\Sigma_{total}(a) = \frac{ \Sigma_{thin}(a) \Sigma_{thick}(a)}
		{ \Sigma_{thin}(a) + \Sigma_{thick}(a)}. 
\end{equation}
After inputting values for $\Sigma_{thin}$ and $\Sigma_{thick}$, Equation~\ref{sigma_total} becomes
\begin{equation} 
	\label{sigma_total2}
	\Sigma_{total}(a) = \frac{\frac{4}{3}\pi n_{p} \sigma_{p} a^3}{1+ \frac{2}{3} n_{p} \sigma_{p} a}
\end{equation}
as the final expression for cross-section dependence as a function of grain size. It is worth stressing that the main purpose of Equation~\ref{sigma_total2} was to describe changes of the cross-section over a broad interval of grain sizes and not to calculate cross-sections to high accuracy. 

With this dependence in hand, theoretically and experimentally calculated cross-sections \citep{Chantler1995, NIST} were applied to Equation~\ref{sigma_total2} to approximate the dust/ice cross-sections over the entire grain radius range. Cometary atmospheric composition was modeled as 85\% H$_{2}$O, 10\% C, and 5\% N and Si based on average cometary composition ratios \citep{Lisse2005, Biver2006, Christian2010}.  Dust and ice densities within the atmosphere were estimated using the empirically established proportionality for mass-loss rates $q_{\rm dust} \simeq 1.5 q_{\rm gas}$, where $q_{\rm gas}$ may be derived from the particle outflow rates $Q_{\rm gas}$ \citep{McDonnell1987, Krasnopolsky2004}. 

Selecting an accurate solar spectrum is crucial as it will dictate the spectral shape of the scattered emissions since we only considered coherent scattering. Ideally, the model would utilize solar spectrum observations taken simultaneously with the comet observations, but those are not available for the selected sample set. The {\it CHIANTI} atomic database was therefore used to model the relative average solar X-ray spectral intensities over the 0.3--3.0 keV energy range \citep{Dere1997, Landi2013} and was normalized with respect to solar intensities over the $4-8\rm\ \AA$ emission wavelength range reported from the \textit{GOES X-Ray Satellite} during the observation time, after adjusting for the differences in travel time. The observed solar intensities are listed in Table~\ref{table:parameter}.

While the CX component has previously been shown to be accurate for cometary spectra below 1 keV (generally agreeing to average cometary spectra within a reduced chi-squared $\chi_{R}^2 < 1.1$; \cite{Snios2016}, J. Lichtman et al. 2018, submitted), the scattering component has several sources of systematic uncertainty, such as the approximate cross-section dependence used. In addition, the scattered spectrum shape is highly dependent on the solar X-ray spectrum, which is known to vary on short time scales. Average solar X-ray spectrum were used for the model, which may present large differences from the observed spectral peak ratios. Comparisons between the total emission spectrum model to observations were therefore focused on the agreement of the total X-ray intensities. Any comparisons between the spectral shape between the {\it ab initio} model and observations should only be treated as indicative.

\section{{\it Chandra} Observations}
\label{sect:chandra}

To test the validity of the emission model, archival cometary data from \chandra{} and \textit{XMM} were analyzed to select prime candidates for comparison (J. Lichtman et al. 2018, submitted). We focused our analysis on detecting the presence of spectral features greater than 1 keV as this energy range was postulated to be where the scattering-to-CX ratio will be greatest. Comets C2012/S1 (ISON) and 153P/Ikeya--Zhang (IZ) were chosen from the available data as they were the only cometary spectra that had spectral features above 1 keV at a signal-to-noise greater than 3. Both comets were observed during periods of high solar X-ray activity, as confirmed by \GOES{}, which may in part explain the high-energy features and above average signal-to-noise ratios. All of the required modeling parameters for each comet are listed in Table~\ref{table:parameter}. 

All of the observations were performed by \chandra{} using the Advanced CCD Imaging Spectrometer (ACIS) with the object centered on the S3 chip in the VFAINT mode. The data were reprocessed using \ciao{} 4.9 with \caldb{} 4.7.4 \citep{Fruscione2006} and were co-added for each comet. On-chip background subtraction was used as it has been shown to be the preferred method for cometary analysis, and the \ciao{} {\it deflare} routine was used to remove background flares. All of the spectra used in the analysis were extracted using the {\it specextract} routine on a circular region of radius $10^5 \rm\ km$ with the comet defined as the center and were subsequently binned to have a minimum of 5 counts per bin. 

\section{Results and Discussion}

\subsection{Modeled Emission Spectrum for Ikeya--Zhang}
\label{sect:IZ}

\begin{figure}
	\includegraphics[width=0.47\textwidth]{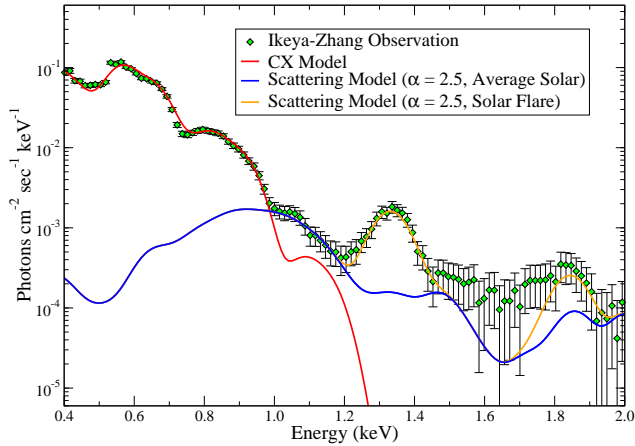}
	\caption{Comparison of the modeled spectral intensity
		contributions from CX and dust/ice particle scattering 
		to the \chandra{} observation of Comet Ikeya--Zhang. 
		The modeled scattering emission includes dust contribution from all grain 
		radii. The scattering model is calculated for both the average solar and 
		solar flare spectrum, with the solar flare spectrum producing an excellent 
		agreement to the observation at energies greater than 1 keV.}	
	\label{fig:IZ_dust}
\end{figure}

The derived total emission model for Comet IZ is compared with the average background-corrected observational \chandra{} spectrum in Figure~\ref{fig:IZ_dust}. CX emission clearly dominates below 1 keV, and the resulting SW composition indicates high-speed SW. This composition indicates the presence of solar flare activity during the observations, which agrees with the \GOES{} observations. In regards to scattering emissions, both the total intensity and spectral shape agree well with the observational data over the 1.0--1.2 keV energy range with an $\alpha=2.5$. Above 1.2 keV, the scattering model begins to diverge from the observed spectrum, albeit at a slower rate than that of the CX model.

Given that the scattered emission features are dependent on the solar spectrum at the time of observation, it is probable that the average solar spectrum used in this model is not a valid approximation of the solar conditions during IZ's perihelion approach based on the difference in spectral shape. In particular, notable discrepancies between the data and model are seen at 1.35 and 1.85 keV which correspond to dust/ice scattering of the resonant Mg XI (2p--1s) and Si XIII (2p--1s)  emission lines present in the solar flare spectrum, respectively. These lines have been seen as elevated during previous periods of solar flare activity similar to those that occurred during the IZ observations \citep{McKenzie1985}. We therefore inferred that using a Solar flare X-ray spectrum may be a more accurate description of the system, so we repeated our analysis with a solar flare spectrum taken from \cite{McKenzie1985}. Using the revised scattering model produced a significantly improved fit to the observation, as shown with the $\chi^2$ results in Table \ref{table:fits}.

Given the notable improvement to the fit from introducing the scattering component to the emission spectrum model, these results provide evidence that the theoretical dust/ice scattering model is able to match observed comet emission intensities over the 1--2 keV energy range. Furthermore, CX remains the dominate emission mechanism below 1 keV by 2--3 orders of magnitude, which is consistent with prior analyses \citep{Krasnopolsky1997, Kharchenko2003, Lisse2004b}. This hybrid model maintains physically consistent solar conditions and SW abundances while also reproducing the observed high-energy spectral features between 1 and 2 keV.

\subsection{Modeled Emission Spectrum for ISON}
\label{sect:ISON}

\begin{figure}
	\includegraphics[width=0.47\textwidth]{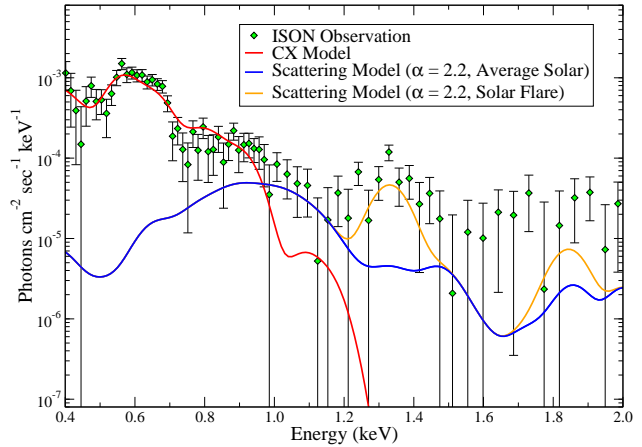}	
	\caption{Comparison of the modeled spectral intensity 
		contributions from CX and dust/ice particle scattering 
		to the \chandra{} observation of Comet ISON. The modeled scattering
		emission includes dust contribution from all grain radii. The 
		scattering model is calculated for both the average solar and solar 
		flare spectrum, with the solar flare spectrum producing a 
		reasonable agreement to observed emission feature at 1.35 keV.}
	\label{fig:ISON_dust}
\end{figure}

A comparison of the theoretical cometary emission model of Comet ISON to its observed spectrum is shown in Figure~\ref{fig:ISON_dust}. CX emissions are again shown to dominate below 1 keV, with the SW composition indicative of high-speed SW outflows (albeit not as high as what was observed for Comet IZ). An agreement in intensity between the observation and the average solar scattered model over the 1.0--1.25 keV energy range was found for $\alpha = 2.2$, as shown in Figure~\ref{fig:ISON_dust}. Although the observation has a lower signal-to-noise than Comet IZ, a clear divergence between the data and model is seen at 1.35 keV, which corresponds to the resonant Mg XI emission line. As Comet ISON was observed during solar flare activity, we again applied a solar flare spectrum to our scattered emission model. The revised model improved the overall $\chi^2$ of the fit, as shown in Table \ref{table:fits}. While poor signal-to-noise makes it impossible to fit any additional features, the reduction in $\chi_{R}^2$ from the inclusion of dust/ice scattering emission to the model suggests that scattering is a viable method for explaining the observed emission features at energies greater than 1 keV.

\begin{table}
	\caption{0.4--2.0 keV Spectrum Model Fit Results}
	\label{table:fits} 
	\begin{tightcenter}
	\begin{tabular}{ c | c c }
		\hline \hline
		& IZ & ISON \\
		Model & ($\chi^2/dof$) & ($\chi^2/dof$) \\
		\hline
		CX & 651.1/102 & 150.7/102 \\
		CX+Dust Scattering\tablenote{Using the average solar 
			spectrum\citep{Lepri2013}} & 443.2/101 & 138.9/101 \\
		CX+Dust Scattering\tablenote{Using the solar flare 
			spectrum \citep{McKenzie1985}} & 274.8/101 & 111.6/101 \\
		\hline
	\end{tabular}
	\end{tightcenter}
\end{table}

\section{Conclusions}
\label{sect:conclusions}

Emissions from cometary atmospheres were modeled from first principles using CX interaction with SW ions as well as coherent scattering from dust and ice grains. Scattering cross-sections were interpolated over the 1 nm--1 cm grain radius range using approximations based on the optically thin or thick nature of the grain, providing a description of the cross-sections over a broad interval of grain sizes. The emission model was compared to \chandra{} observations of Comets ISON and IZ, which were selected due to their high signal-to-noise ratios and clear presence of spectral features between 1 and 2 keV. Comparing the observations to the theoretical models showed that CX is the dominate emission mechanism below 1 keV, with both comets showing evidence of high-speed SW outflows. Inclusion of scattered emissions to the models produced notable improvements in the model fits to the data, indicating that scattering is a significant emission mechanism in cometary systems. Scattering was also shown to be the dominate mechanism over the 1--2 keV range with both spectral features and intensities between the model and data trending similarly, a surprising outcome given the lack of precise solar X-ray spectra required for a more accurate fit. Varying the atmospheric dust/ice density dependence with respect to grain size as $n(a) \propto a^{-\alpha}$ was required to equal the observed spectral intensities, with IZ requiring $\alpha = 2.5$ and ISON requiring $\alpha = 2.2$. The results are both physically consistent and agree with independent observations and simulations, indicating that the dust/ice density dependence on the grain radius varies significantly between comets. These results provide evidence that the theoretical dust/ice scattering model is a significant emission mechanism in cometary systems at energies greater than 1 keV, particularly during periods of solar flare activity.

Further improvements, such as the introduction of accurate grain morphology in the cross-section analysis or utilizing an accurate solar spectrum taken simultaneously with the comet observations, should be incorporated into the scattering model to more accurately quantify dust/ice densities as well as grain size dependence. Observations of comets made during close perigees, such as Comet 46P/Wirtanen, which will come within 0.08 au of Earth in 2018 December, will also improve the accuracy of the emission model as we may assume that the physical conditions of the comet are similar to the results from Earth-orbiting satellites rather than extrapolating these values from average results, solar activity models, and time of flight corrections, all of which add systematic uncertainty. However, the best way to further these results is with additional observations, as a larger sample size of comets with clear high-energy features is required before rigorous, quantitative conclusions on dust and ice distributions in cometary systems can be made. 

\acknowledgements{
The scientific results reported in this article are based in part on data obtained from the \chandra{} Data Archive. We also acknowledge the National Oceanic and Atmospheric Administration for their \GOES{} X-ray data.
}


\bibliographystyle{aasjournal}
\bibliography{all_data}

\end{document}